\documentclass{algotel}

\usepackage{tikz}
\usepackage{pgfplots}
\usepackage[noabbrev]{cleveref}
\usepackage{algorithm}
\usepackage{algpseudocodex}

\floatname{algorithm}{Algorithme}

\usepackage[colorinlistoftodos,textsize=tiny]{todonotes}

\title{\vspace{-.3cm}Trier des cochons sauvages\thanks{Ce travail a été effectué au LINCS (\url{https://www.lincs.fr/}).}}

\author{%
Emma Caizergues\addressmark{1, 2},
François Durand\addressmark{2}
\and Fabien Mathieu\addressmark{3}%
}

\address{%
\addressmark{1}Université Paris Dauphine, France\\
\addressmark{2}Nokia, Massy, France\\
\addressmark{3}Swapcard, Paris, France%
}


\begin{document}
\maketitle

\begin{abstract}
Chjara, éleveuse à Cargèse, possède $n$ cochons sauvages. Elle aimerait trier son troupeau par poids afin de mieux satisfaire les demandes de ses acheteurs et acheteuses. 
Chaque bête a un poids distinct, hélas inconnu de Chjara. 
Elle n'a à sa disposition qu'une balance de Roberval, qui ne lui permet de comparer deux porcins qu'au prix d'une man\oe uvre acrobatique. La balance, assez vétuste, peut se briser à tout moment. Chjara veut donc trier son troupeau en un minimum de pesées, mais aussi avoir une bonne estimation du résultat après chaque pesée.

Pour aider Chjara, nous posons le problème de la recherche d'un bon \emph{tri interruptible}, au sens de la distance tau de Kendall entre résultat provisoire et liste parfaitement triée, et nous apportons les contributions suivantes :
\begin{itemize}
	\item Nous introduisons \emph{Corsort}, une famille de tris interruptibles reposant sur des estimateurs.
	\item Par simulation, nous montrons qu'un Corsort bien configuré a un temps de terminaison quasi-optimal, et fournit de meilleures estimations intermédiaires que les meilleurs tris dont nous avons connaissance.
\end{itemize}

\end{abstract}





\vspace{-.3cm}
\section{Contexte}\label{sec_contexte}

Pour trier des cochons sauvages, comparer est beaucoup plus coûteux que toute autre opération. Des problèmes similaires apparaissent si la comparaison implique une intervention humaine ou des données massives. On doit alors séparer la complexité en comparaisons de la complexité globale~\cite{giesen2009approximate}.

Si la balance de Chjara était indestructible, on retrouverait un problème classique : minimiser le nombre de comparaisons \cite{cormen}. En plus d'algorithmes célèbres tels que le tri rapide, le tri fusion et le tri par tas (parmi bien d'autres), on peut citer l'algorithme de Ford-Johnson \cite{Ford1959ATP}, extrêmement proche de la borne théorique en nombre de comparaisons, et même optimal pour certaines valeurs~\cite{peczarski2004new}.


La fragilité de la balance nous rapproche des \emph{algorithmes interruptibles} (\emph{anytime algorithms}), qui maintiennent à tout instant une estimation du résultat~\cite{zilberstein1996using}. Étonnamment, les tris ont été peu étudiés dans cette littérature : les études ne concernent que le tri par sélection, le tri de Shell ou le tri rapide, sans introduire d'algorithme plus adapté, et les mesures d'écart au résultat final ne sont pas des distances \cite{horvitz1988reasoning,grass1995programming}.


Parmi les problèmes connexes, les \emph{algorithmes progressifs} peuvent aussi être interrompus à tout moment, mais l'accent est mis sur des bornes prouvables de performance en pire cas plutôt que sur l'efficacité moyenne empirique~\cite{alewijnse2014progressive}. Les \emph{algorithmes par contrat} opèrent également un compromis entre temps et précision, mais supposent que le temps disponible est connu à l'avance~\cite{zilberstein1996using}. À l'inverse, le \emph{tri approximatif} part d'un objectif en terme d'erreur maximale, et essaie de borner le nombre d'opérations nécessaire pour l'atteindre. L'algorithme ASort donne des garanties de cette nature \cite{giesen2009approximate}.

Dans la suite de cet article, 
la \cref{sec_estimateurs} présente les notions de tri interruptible, d'estimateur et de tri Corsort; la \cref{sec_eval} évalue la qualité des solutions considérées par des simulations; et la \cref{sec_conclu} conclut.

\section{Tris interruptibles}\label{sec_estimateurs}

Formellement, nous voulons trier une liste $X = (X[1], \ldots, X[n])$, où $n > 0$, en effectuant des comparaisons du type: \emph{est-ce que $X[i] < X[j]$ ?} Un \emph{tri interruptible} (\emph{anytime sorting algorithm}) est un algorithme capable, à chaque étape~$k$ de son exécution, de renvoyer une estimation $X_k$ du résultat. Dans notre modèle, chaque comparaison constitue une étape de l'algorithme\footnote{Par convention, si l'algorithme termine en moins de $k$ comparaisons, alors $X_k$ est le résultat final, c'est-à-dire la liste triée.}, et on mesure la qualité de $X_k$ par la distance tau de Kendall~\cite{kendall1938measure} entre $X_k$ et la liste triée: $\tau(X_k) = |\{ (i, j) : i < j, X_k[i] > X_k[j] \}|$. 
Idéalement, nous cherchons un tri interruptible  dont le \emph{profil de performance} $k \to \tau(X_k)$, représentant l'erreur commise, est constamment plus bas que celui des autres algorithmes testés.

\subsection{Tris classiques}

Certains algorithmes classiques peuvent être vus comme interruptibles car ils maintiennent une liste courante qui converge vers la liste triée et peut servir d'estimation $X_k$. C'est le cas du tri rapide et du tri fusion, que nous avons implantés d'une manière favorable à l'esprit de l'algorithme initial: par exemple, pour le tri rapide, la position du pivot est mise à jour dans la liste après chaque comparaison.



Modifier l'ordre des comparaisons effectuées peut améliorer les estimations intermédiaires $X_k$. Pour le tri fusion, naturellement on parcourt l'arbre de récursion en profondeur (DFS), mais on peut aussi le parcourir en largeur \footnote{Dans le cas simple où la taille de la liste est une puissance de 2, on fusionne tous les sous-tableaux de taille $1$, puis $2$, puis $4$, etc.} (BFS). Pour le tri rapide, notre implémentation améliorée est équivalente à l'algorithme ASort \cite{giesen2009approximate}, en utilisant la sélection rapide \cite{hoare1961algorithm} comme sous-algorithme d'identification de la médiane.



D'autres algorithmes classiques permettent d'obtenir une estimation $X_k$ par une transformation simple de l'état courant. C'est le cas du tri par tas: parcourir le tas à l'envers, puis les éléments déjà triés à l'endroit. 

Enfin, certains algorithmes comme Ford-Johnson n'ont pas d'estimation respectant <<~l'esprit~>> de l'algorithme. La section suivante montre comment produire des $X_k$ pour n'importe quel algorithme de tri.







%

\subsection{Tris classiques avec estimateurs}



Pour rendre n'importe quel tri interruptible, nous proposons de construire un estimateur qui ignore l'algorithme de tri utilisé et repose uniquement sur l'historique des comparaisons effectuées.


Notons $C_k=\{X[i_1]<X[j_1], \ldots, X[i_k]< X[j_k]\}$ le résultat de $k$ comparaisons. $C_k$ définit par clôture transitive un ordre partiel $\preceq_k$ sur les indices de la liste initiale (et donc ses éléments). Un \emph{estimateur} est une fonction qui associe à tout ordre partiel un ordre total compatible.


Une première idée est de considérer l'ensemble des ordres totaux compatibles avec $\preceq_k$ (ses \emph{extensions linéaires}), et d'associer à chaque élément un score correspondant à sa hauteur moyenne dans les extensions linéaires de $\preceq_k$. On renvoie ensuite la liste issue du tri des scores (trier $n$ scores est bien moins coûteux que comparer deux cochons sauvages). Cet estimateur semble convainquant mais son coût est prohibitif : compter l'ensemble des extensions linéaires d'un ordre partiel est déjà \#P-complet \cite{brightwell1991counting}.


On propose donc une fonction de score 
heuristique pour calculer en $O(n^2)$ une estimation $X_k$ raisonnable. Si $i$ est un indice de la liste initiale, on note $d_k(i)=|\{j\in [n]: j \preceq_k i\}|$ et $a_k(i)=|\{j\in [n]: i \preceq_k j\}|$ le nombre de descendants et d'ancêtres de $i$ (lequel est inclus dans les deux ensembles par convention). On définit la fonction de score $\rho_k$ par $\rho_k(i)=d_k(i)/(d_k(i)+a_k(i))$. Cela revient à positionner $i$ comme si ses descendants et ses ancêtres avaient en moyenne des positions régulièrement espacées.



\subsection{Tris orientés comparaisons (Corsort)}\label{sec_corsort}

Un tri \emph{Corsort} (\emph{Comparison-ORiented Sort}) fonctionne ainsi : en fonction de l'ordre partiel courant $\preceq_k$, estimer $X_k$ et choisir la comparaison suivante. On suppose qu'on choisit toujours des paires non comparables selon $\preceq_k$, ce qui assure de terminer en au plus $n (n -1) / 2$ comparaisons.


Comme pour les tris classiques, nous utiliserons la fonction de score $\rho$ pour construire l'estimateur.



Pour la fonction de prochaine comparaison, on doit assurer, à long terme, une terminaison rapide. C'est un problème de \emph{tri sous information partielle}, qui revient à choisir une comparaison dont les deux issues sont aussi équiprobables que possible~\cite{supi}. À cette fin, nous définissons la fonction de score $\Delta_k$ par $\Delta_k(i)=d_k(i)-a_k(i)$. $\Delta_k$ attribue à chaque $i$ un score qui reflète la moyenne entre sa plus basse et sa plus haute positions possibles, et nous souhaitons comparer des éléments dont les scores sont proches.
À court terme, pour améliorer $X_k$, il faut acquérir de l'information sur les éléments pour lesquels on en a peu. On introduit donc $I_k(i) = a_k(i) + d_k(i)$, et on souhaite comparer des éléments pour lesquels $I_k$ est faible.


Après moult expérimentations, parmi les paires encore incomparables, nous choisissons la paire $(i, j)$ qui minimise lexicographiquement la paire $\big(|\Delta_k(i)-\Delta_k(j)|, \max(I_k(i), I_k(j))\big)$.


\section{Évaluation}\label{sec_eval}

Nous avons développé un paquet Python pour créer des tris interruptibles et mesurer leurs performances\footnote{\url{https://emczg.github.io/corsort/}}. Pour $n$ donné, nous tirons 10\,000 permutations aléatoires et calculons les comparaisons nécessaires pour un tri complet et les profils de performance $k \to \tau(X_k)$. Pour donner un aperçu de la distribution des résultats, nous traçons pour chaque algorithme la médiane (courbe foncée), les quantiles de 25\% à 75\% (zone claire), et les quantiles de 2,5\% à 97,5\% (zone très claire) qui représentent un intervalle de confiance à 95\%.


La Figure \ref{fig:total} montre le temps de terminaison (en comparaisons) pour des valeurs de $n$ allant de $8$ à $1024$ et les tris suivants : par tas, rapide, Corsort, fusion et Ford-Johnson. L'axe des ordonnées montre l'écart relatif par rapport à la borne inférieure $n \log_2(n) - n/\ln(2) + \log_2(2\pi  n)/2$~\cite{cormen} : plus une courbe est proche de $0$, plus elle est optimale.
\begin{figure}[ht]
	\centering
\begin{tikzpicture}

\definecolor{lightgray204}{RGB}{204,204,204}
\definecolor{darkgray176}{RGB}{176,176,176}

\definecolor{marronEm}{RGB}{152,78,46} 
\definecolor{blackEm}{RGB}{15,15,15}

\definecolor{redEm}{RGB}{214,39,40}

\definecolor{purpleEm}{RGB}{148,103,189} 
\definecolor{blueEm}{RGB}{31,119,180}

\definecolor{greenEm}{RGB}{44,160,44}

\definecolor{orangeEm}{RGB}{246,132,18}


\begin{axis}[width=15cm, height=8cm,
legend cell align={left},
legend style={
  fill opacity=0.8,
  draw opacity=1,
  text opacity=1,
  at={(0.91,0.91)},
  anchor=north east,
  draw=lightgray204
},
log basis x={10},
tick align=outside,
x grid style={darkgray176},
xlabel={Taille $n$ du troupeau},
xmin=8, xmax=1024,
xmode=log,
xtick pos=left,
xtick style={color=black},
y grid style={darkgray176},
ylabel={Écart à la borne théorique (\%)},
ymin=0, ymax=100,
ytick pos=both,
ytick style={color=black}
]
\path [draw=orangeEm, fill=orangeEm, opacity=0.4]
(axis cs:8,76.6531559510697)
--(axis cs:8,63.5677369917312)
--(axis cs:16,78.5608238565605)
--(axis cs:32,86.9802161041612)
--(axis cs:64,91.5584123652679)
--(axis cs:128,94.2301272769286)
--(axis cs:256,95.7249516630602)
--(axis cs:512,96.6625487004642)
--(axis cs:1024,97.2629503084852)
--(axis cs:1024,97.7191023498968)
--(axis cs:1024,97.7191023498968)
--(axis cs:512,97.3850984135745)
--(axis cs:256,96.9126030687827)
--(axis cs:128,96.0453621112924)
--(axis cs:64,94.5990220853515)
--(axis cs:32,92.0796765433656)
--(axis cs:16,87.6018782290445)
--(axis cs:8,76.6531559510697)
--cycle;

\path [draw=orangeEm, fill=orangeEm, opacity=0.2]
(axis cs:8,83.195865430739)
--(axis cs:8,50.4823180323927)
--(axis cs:16,69.5197694840764)
--(axis cs:32,81.8807556649569)
--(axis cs:64,88.5178026451842)
--(axis cs:128,92.2752589937676)
--(axis cs:256,94.5966828276239)
--(axis cs:512,95.939998987354)
--(axis cs:1024,96.8067982670735)
--(axis cs:1024,98.1296391871674)
--(axis cs:1024,98.1296391871674)
--(axis cs:512,98.0560374328911)
--(axis cs:256,97.9221067636468)
--(axis cs:128,97.720963496859)
--(axis cs:64,97.3017862809814)
--(axis cs:32,96.3292269093693)
--(axis cs:16,92.1224054152866)
--(axis cs:8,83.195865430739)
--cycle;

\path [draw=purpleEm, fill=purpleEm, opacity=0.4]
(axis cs:8,24.3114801137157)
--(axis cs:8,-1.85935780496127)
--(axis cs:16,1.71186169044584)
--(axis cs:32,8.78848936969381)
--(axis cs:64,12.5025596430938)
--(axis cs:128,16.4542962968788)
--(axis cs:256,19.121435993962)
--(axis cs:512,21.6722106183819)
--(axis cs:1024,23.3663195997915)
--(axis cs:1024,32.7630516528723)
--(axis cs:1024,32.7630516528723)
--(axis cs:512,32.2782081929641)
--(axis cs:256,31.4730106134755)
--(axis cs:128,30.4176411766005)
--(axis cs:64,28.7191448168731)
--(axis cs:32,26.6366009069092)
--(axis cs:16,24.314497621656)
--(axis cs:8,24.3114801137157)
--cycle;

\path [draw=purpleEm, fill=purpleEm, opacity=0.2]
(axis cs:8,57.025027512062)
--(axis cs:8,-14.9447767642998)
--(axis cs:16,-7.32919268203824)
--(axis cs:32,-1.41043150871498)
--(axis cs:64,3.7185760072967)
--(axis cs:128,8.35555626664026)
--(axis cs:256,11.5204669973383)
--(axis cs:512,14.9628204252156)
--(axis cs:1024,17.4363430614395)
--(axis cs:1024,46.4932280993641)
--(axis cs:1024,46.4932280993641)
--(axis cs:512,47.1162826586205)
--(axis cs:256,48.0407477233037)
--(axis cs:128,49.9663240082109)
--(axis cs:64,51.6926404797271)
--(axis cs:32,53.8337232493327)
--(axis cs:16,55.9581879253503)
--(axis cs:8,57.025027512062)
--cycle;

\path [draw=greenEm, fill=greenEm, opacity=0.4]
(axis cs:8,4.68335167470799)
--(axis cs:8,-1.85935780496127)
--(axis cs:16,-0.548401902675189)
--(axis cs:32,1.13929871088723)
--(axis cs:64,2.02934838502804)
--(axis cs:128,3.18911866114324)
--(axis cs:256,3.97888057100073)
--(axis cs:512,4.61487631960136)
--(axis cs:1024,5.12023794332386)
--(axis cs:1024,5.65621659198259)
--(axis cs:1024,5.65621659198259)
--(axis cs:512,5.38903672650517)
--(axis cs:256,5.10714940643706)
--(axis cs:128,4.86472004670984)
--(axis cs:64,4.39426705620418)
--(axis cs:32,4.53893900369016)
--(axis cs:16,3.97212528356685)
--(axis cs:8,4.68335167470799)
--cycle;

\path [draw=greenEm, fill=greenEm, opacity=0.2]
(axis cs:8,11.2260611543772)
--(axis cs:8,-8.40206728463052)
--(axis cs:16,-5.06892908891723)
--(axis cs:32,-2.26034158191571)
--(axis cs:64,0.00227523830562681)
--(axis cs:128,1.51351727557663)
--(axis cs:256,2.90999430585053)
--(axis cs:512,3.84071591269755)
--(axis cs:1024,4.60706689673571)
--(axis cs:1024,6.16938763857076)
--(axis cs:1024,6.16938763857076)
--(axis cs:512,6.13739178651218)
--(axis cs:256,6.17603567158729)
--(axis cs:128,6.40068798347921)
--(axis cs:64,6.75918572738032)
--(axis cs:32,7.08866922329234)
--(axis cs:16,8.49265246980888)
--(axis cs:8,11.2260611543772)
--cycle;

\path [draw=blackEm, fill=blackEm, opacity=0.4]
(axis cs:8,11.2260611543772)
--(axis cs:8,-1.85935780496127)
--(axis cs:16,1.71186169044584)
--(axis cs:32,1.98920878408795)
--(axis cs:64,2.02934838502804)
--(axis cs:128,2.21168451956273)
--(axis cs:256,2.13802089213093)
--(axis cs:512,2.034341629922)
--(axis cs:1024,1.89296225033615)
--(axis cs:1024,2.16665347518317)
--(axis cs:1024,2.16665347518317)
--(axis cs:512,2.47303252716748)
--(axis cs:256,2.8506117355644)
--(axis cs:128,3.32875210994046)
--(axis cs:64,4.05642153175043)
--(axis cs:32,5.38884907689088)
--(axis cs:16,6.23238887668787)
--(axis cs:8,11.2260611543772)
--cycle;

\path [draw=blackEm, fill=blackEm, opacity=0.2]
(axis cs:8,11.2260611543772)
--(axis cs:8,-14.9447767642998)
--(axis cs:16,-5.06892908891723)
--(axis cs:32,-1.41043150871498)
--(axis cs:64,0.00227523830562681)
--(axis cs:128,1.09461692918498)
--(axis cs:256,1.42543004869746)
--(axis cs:512,1.62145607957329)
--(axis cs:1024,1.63067482652444)
--(axis cs:1024,2.40613329692432)
--(axis cs:1024,2.40613329692432)
--(axis cs:512,2.86011273061939)
--(axis cs:256,3.50382000871177)
--(axis cs:128,4.30618625152097)
--(axis cs:64,5.74564915401912)
--(axis cs:32,7.08866922329234)
--(axis cs:16,10.7529160629299)
--(axis cs:8,11.2260611543772)
--cycle;

\path [draw=redEm, fill=redEm, opacity=0.4]
(axis cs:8,4.68335167470799)
--(axis cs:8,-1.85935780496127)
--(axis cs:16,-0.548401902675189)
--(axis cs:32,0.289388637686483)
--(axis cs:64,0.00227523830562681)
--(axis cs:128,0.25681623640168)
--(axis cs:256,0.237778642975006)
--(axis cs:512,0.279578040940032)
--(axis cs:1024,0.262218702289374)
--(axis cs:1024,0.399064314712883)
--(axis cs:1024,0.399064314712883)
--(axis cs:512,0.486020816114374)
--(axis cs:256,0.594074064691741)
--(axis cs:128,0.815350031590545)
--(axis cs:64,1.01581181166683)
--(axis cs:32,1.98920878408795)
--(axis cs:16,1.71186169044584)
--(axis cs:8,4.68335167470799)
--cycle;

\path [draw=redEm, fill=redEm, opacity=0.2]
(axis cs:8,4.68335167470799)
--(axis cs:8,-8.40206728463052)
--(axis cs:16,-5.06892908891723)
--(axis cs:32,-2.26034158191571)
--(axis cs:64,-1.01126133505559)
--(axis cs:128,-0.301717558787185)
--(axis cs:256,-0.0591342084556024)
--(axis cs:512,0.0731352657656892)
--(axis cs:1024,0.136776890901147)
--(axis cs:1024,0.513102325065806)
--(axis cs:1024,0.513102325065806)
--(axis cs:512,0.666658244391938)
--(axis cs:256,0.890986916122372)
--(axis cs:128,1.23425037798219)
--(axis cs:64,1.69150286057429)
--(axis cs:32,2.83911885728869)
--(axis cs:16,3.97212528356685)
--(axis cs:8,4.68335167470799)
--cycle;

\addplot [semithick, orangeEm, mark=*, mark size=1, mark options={solid}]
table {%
8 70.1104464714005
16 83.0813510428025
32 89.5299463237635
64 93.2476399875365
128 95.2075614185091
256 96.3187773659214
512 97.0238235570193
1024 97.491026329191
};
\addlegendentry{Tas}
\addplot [semithick, purpleEm, mark=*, mark size=1, mark options={solid}]
table {%
8 4.68335167470799
16 10.7529160629299
32 16.4376800285004
64 19.5973156566223
128 22.5981680439564
256 24.4658673197131
512 26.420394447392
1024 27.6541487890613
};
\addlegendentry{Rapide}
\addplot [semithick, greenEm, mark=*, mark size=1, mark options={solid}]
table {%
8 4.68335167470799
16 1.71186169044584
32 2.83911885728869
64 3.38073048284298
128 4.02691935392654
256 4.51332370357584
512 5.00195652305326
1024 5.38252536713559
};
\addlegendentry{Corsort}
\addplot [semithick, blackEm, mark=*, mark size=1, mark options={solid}]
table {%
8 4.68335167470799
16 3.97212528356685
32 3.68902893048941
64 3.04288495838925
128 2.77021831475159
256 2.49431631384767
512 2.26658975199314
1024 2.02980786275966
};
\addlegendentry{Fusion}
\addplot [semithick, redEm, mark=*, mark size=1, mark options={solid}]
table {%
8 4.68335167470799
16 1.71186169044584
32 1.13929871088723
64 0.677966287213083
128 0.536083133996113
256 0.415926353833385
512 0.382799428527214
1024 0.330641508501128
};
\addlegendentry{Ford-Johnson}
\end{axis}

\end{tikzpicture}
	\vspace{-1.2cm}
	\caption{Nombre relatif de comparaisons supplémentaires par rapport à la borne théorique.}
	\label{fig:total}
\end{figure}
Notons que le tri par tas compare presque deux fois plus que nécessaire. Le tri rapide est meilleur (moins de 30\% de surcoût)
mais a une grande variance. Les trois tris restants ont un surcoût encore plus faible (5\% pour Corsort, 2\% pour le tri fusion, 0,3\% pour Ford-Johnson) et une variance négligeable. Nous concluons que Corsort est un bon candidat puisqu'il n'est battu que par des algorithmes dont la terminaison est asymptotiquement optimale, i.e.  équivalente à $n\log_2(n)$~\cite{Ford1959ATP}.



La Figure \ref{fig:traj} montre les profils de performance des quatre meilleurs tris de la Figure \ref{fig:total} (rapide, Corsort, fusion, et Ford-Johnson). Pour les estimateurs, Corsort et Ford-Johnson utilisent $\rho$. Pour les tris rapide et fusion deux variantes sont utilisées : la version de base avec son estimation naturelle (l'état courant de la liste) ; une version améliorée (respectivement fusion-BFS et ASort) munie d'un estimateur utilisant~$\rho$. La métrique utilisée est la distance à la liste triée: plus elle est faible, meilleure est la performance.
\begin{figure}[ht]
	\centering
	\include{trajectories}
	\vspace{-1.2cm}
	\caption{Profils de performance de tris interruptibles pour $n=1000$. La distance $\tau$ est normalisée par $n(n-1)/2$.}
	\label{fig:traj}
\end{figure}
La principale observation est la supériorité du profil de Corsort : il est monotone et constamment sous les autres à part en terminaison (il termine un peu plus tard que fusion ou Ford-Johnson). Corsort est donc un excellent tri interruptible que nous recommandons à Chjara pour trier ses cochons sauvages. On remarque aussi que l'utilisation de l'estimateur $\rho$ améliore le profil de performance\footnote{Non montré par manque de place, $\Delta$ améliore également le profil mais est généralement moins performant que $\rho$.}, parfois au prix d'un comportement non-monotone. Enfin, on peut constater la relative bonne performance du tri fusion-BFS muni de $\rho$, qui peut être un choix intéressant si l'on désire un tri interruptible qui soit également rapide, avec une terminaison en $O(n\log(n))$ opérations totales (pas seulement en comparaisons) hors estimateur. 


\section{Conclusion}\label{sec_conclu}

Nous avons étudié des tris interruptibles générant un minimum de comparaisons. Nous avons proposé une méthode pour rendre tout tri interruptible, avec interruption possible après chaque comparaison. Nous avons introduit Corsort, une famille de tris à base d'estimateurs. Par simulation, nous avons montré qu'un tri Corsort bien configuré a un temps de terminaison (en nombre de comparaisons) quasi-optimal et possède un profil de performance meilleur que les meilleurs tris dont nous avons connaissance.


\bibliographystyle{alpha}
\bibliography{corsort}
\label{sec:biblio}

\end{document}